\documentclass[twocolumn,3p]{elsarticle}

\usepackage[T1]{fontenc}
\usepackage{hyperref}
\usepackage{xcolor}

\usepackage[type={CC},modifier={by-nc-nd},version={4.0},hyperxmp={false}]{doclicense}

\date{}

\journal{Information and Software Technology}

\usepackage{array}
\usepackage{textgreek}
\usepackage{prettyref}
\newrefformat{guide}{G\ref{#1}}
\usepackage{multirow}

\begin{document}

\begin{frontmatter}

\title{HyMap: eliciting hypotheses in early-stage software startups using cognitive mapping\tnoteref{mytitlenote}}
\tnotetext[mytitlenote]{This document is the accepted manuscript of the following paper (please cite as this):\\ J. Melegati, E. Guerra, X. Wang. ``HyMap: eliciting hypotheses in early-stage software startups using cognitive mapping.'' Information and Software Technology, vol. 144, p. 106807, Apr. 2022, doi: \url{https://doi.org/10.1016/j.infsof.2021.106807}.  \doclicenseThis}

\author[mymainaddress]{Jorge Melegati\corref{mycorrespondingauthor}}
\cortext[mycorrespondingauthor]{Corresponding author}
\ead{jorge.melegati@unibz.it}

\author[mymainaddress]{Eduardo Guerra}
\ead{eduardo.guerra@unibz.it}

\author[mymainaddress]{Xiaofeng Wang}
\ead{xiaofeng.wang@unibz.it}

\address[mymainaddress]{Free University of Bozen-Bolzano, Piazza Domenicani 3, Bolzano, Italy}

\begin{abstract}
	\textbf{Context:} Software startups develop innovative, software-intensive products. Given the uncertainty associated with such an innovative context, experimentation, an approach based on validating assumptions about the software product through data obtained from diverse techniques, like A/B tests or interviews, is valuable for these companies. Relying on data rather than opinions reduces the chance of developing unnecessary products or features, improving the likelihood of success, especially in early development stages, when implementing unnecessary features represents a higher risk for companies' survival.  Nevertheless, researchers have argued that the lack of clearly defined practices led to limited adoption of experimentation. Since the first step of the approach is to define hypotheses, testable statements about the software product features, based on which software development teams will create experiments, eliciting hypotheses is a natural first step to develop practices.
	\textbf{Objective:} We aim to develop a systematic technique for identifying hypotheses in early-stage software startups to support experimentation in these companies and, consequently, improve their software products.
	\textbf{Methods:} We followed a Design Science approach consisting of an artifact construction process, divided in three phases, and an evaluation within three startups.
	\textbf{Results:} We developed the HyMap, a hypotheses elicitation technique based on cognitive mapping. It consists of a process conducted by a facilitator using pre-defined questions, supported by a visual language to depict a cognitive map representing the founder's understanding of the product. Our evaluation showed that founders perceived the artifacts as clear, easy to use, and useful leading to hypotheses and facilitating their idea's visualization. 
	\textbf{Conclusion:} From a theoretical perspective, our study provides a better understanding of the guidance founders use to develop their startups and, from a practical point of view, a technique to identify hypotheses in early-stage software startups.
\end{abstract}


\end{frontmatter}

\section{Introduction}

``Good sense is, of all things among men, the most equally distributed''~\cite{Descartes2020}. With this phrase, Descartes begins his \textit{Discourse on the method} that changed science and human history. In that book, the philosopher introduces the scientific method, a systematic way to derive knowledge based on experiments. Recently, a new trend strengthens this idea of systematically deriving knowledge rather than relying on personal intuition about products in software engineering: experimentation~\cite{Lindgren2016,Fabijan2017}. This approach is a process of continuously validating product assumptions, transforming them into hypotheses, prioritizing, and applying the scientific method to test these hypotheses, supporting or refuting them~\cite{Lindgren2016}. In this context, practitioners can employ several techniques like iterations with prototypes, gradual rollouts, and controlled experiments~\cite{Fabijan2018} but also problem and solution interviews~\cite{Lindgren2016}. 

In a recent position paper~\cite{Melegati2019}, we compared different models of experimentation, like HYPEX~\cite{Olsson2014} and RIGHT~\cite{Fagerholm2017}. We observed that these models follow common steps. First, software development teams identify, specify, and prioritize hypotheses. Based on these hypotheses, they design and execute experiments and, finally, analyze the data collected to confirm or refute the hypotheses. Drawing a parallel to Requirements Engineering activities employed in a requirement-driven approach, we call the step to identify, specify, and prioritize hypotheses as Hypotheses Engineering and argued the need for more systematic practices for these activities.

Given the confusion between the terms ``assumption'' and ``hypothesis,'' especially among practitioners that often use them interchangeably~\cite{Melegati2021}, it is fundamental to differentiate them. Throughout this paper, ``assumption'' refers to a personal or team-wise, generally implicit, understanding taken as truth without being questioned or proved. Meanwhile, a ``hypothesis'' is an explicit statement that has not been proved yet but could be tested through an experiment. In short, assumptions are cognitive and abstract ideas that could be transformed into hypotheses~\cite{Lindgren2016}, concrete elements employed in experimentation. For instance, a product owner believes that the users of an e-commerce website want to save products to be bought later. Based on this assumption, the team develops the functionality. If the product owner instead wants to test if the assumption is valid, a hypothesis statement could be created: ``the users want to save products to be bought later.'' The team, then, could use this hypothesis to create an experiment to assess it, for example, by showing some users a button providing a new function, to gauge their interest in that function, but without implementing the feature.

In this paper, we target hypotheses engineering in the context of software startups. Software startups are organizations looking for a sustainable business model for an innovative product or service they develop where software is a core element~\cite{Unterkalmsteiner2016}. Although we can quickly identify several successful stories like Airbnb or Uber, most startups fail~\cite{Herrmann2012}. Reasons for the lack of success are various: demanding market conditions, lack of team commitment, financial issues~\cite{Klotins2018}, including an inaccurate business development~\cite{Cantamessa2018}, that is, the creation of a non-viable business model. Since a defining characteristic of software startups is the development of an innovative solution, experimentation is a critical element in this context~\cite{Kerr2014}. The value of experimentation is corroborated by the ideas of Lean Startup~\cite{Frederiksen2017,Bortolini2018}, the most well-known methodology among practitioners, which places a strong emphasis on experimentation. 

Nevertheless, startups still tend to focus on developing their proposed solution instead of focusing on the necessary learning process~\cite{Giardino2014,Gutbrod2017}. This aspect is essential, especially in early-stage startups for which developing wrong features may represent a total consumption of the limited financial and human resources available to the company and lead to the company ending. One of the reasons for this limited adoption of experimentation is the lack of clearly defined practices~\cite{Lindgren2016}. Therefore, an essential step for a better implementation of experimentation is a systematic way to specify and handle hypotheses~\cite{Melegati2019}. In this study, our goal is to develop a novel technique to identify the hypotheses on which early-stage software startups base their products. Based on hypotheses, these companies could perform experiments and progress with more precise information about the user and market needs. Therefore, to guide this study, we proposed the following research question: 

\begin{center}
	\textbf{RQ: How can hypotheses be systematically defined in early-stage software startups to support experimentation?}
\end{center} 

To achieve our goal, we followed a Design Science Research (DSR) approach based on Hevner et al.'s guidelines~\cite{Hevner2004} consisting of the artifact construction and evaluation. The artifact construction step is composed of three phases. The first phase's goal was to understand how the assumptions on which startups base their products are formed. In the second and third phases, we proposed, assessed, and improved HyMap, a technique to elicit hypotheses based on cognitive mapping systematically created through a set of questions. We evaluated the practice using a multiple-case study with three software startups. The results indicate that founders perceive the technique as clear, easy to use, self-contained, and useful leading to hypotheses of three types: problem, value, and product. This paper extends a previous paper~\cite{Melegati2020} that presented the first two phases of this study. This study contributes both to practice, providing a novel hypotheses elicitation technique to early-stage software startups, and to theory, by describing how founders create their ideas based on their previous experience and how the steps in this process relate to different types of hypotheses.

The remaining of this paper is organized as follows: Section~\ref{sec:literature_review} presents the background and related work, including the justificatory knowledge~\cite{Hevner2013,Peffers2018} to support the artifact's effectiveness. Section~\ref{sec:research_method} presents the DSR method and how we performed the artifact development process and evaluation. Section~\ref{sec:development_process} details the intermediate results of the artifact development process, Section~\ref{sec:artifact_description} describes the final artifact, and Section~\ref{sec:evalution} presents its evaluation. In Section~\ref{sec:discussion}, we discuss the results and, finally, Section~\ref{sec:conclusions} concludes the paper.

\section{Background and Related Work}
\label{sec:literature_review}

Gregor and Hevner~\cite{Hevner2013} made a distinction between descriptive knowledge (\textOmega) and prescriptive knowledge (\textLambda). While descriptive knowledge concerns the ``what'' about phenomena, including laws and theories to describe natural, artificial, or human phenomena, the prescriptive knowledge focuses on the ``how'' of human-built artifacts, including constructs, models, and methods. According to Gregor and Hevner, in DSR, it is important to review both descriptive and prescriptive knowledge to avoid the lack of novelty and, consequently, a potential lack of contribution to knowledge. Besides that, this literature review should include justificatory knowledge, that is, elements used to inform the artifact construction and explain its effectiveness~\cite{Hevner2013,Peffers2018}.

We organized this section to fulfill the above requirements but also to present key concepts for this study. Section~\ref{ssec:software_startups} presents concepts related to software startups and their lifecycle, and Section~\ref{ssec:experimentation} displays the process of experimentation in software engineering, including hypotheses engineering. In Section~\ref{ssec:available_solutions}, we present available solutions from the scientific and practitioner-produced literature and argue the need for our study. Finally, Section~\ref{ssec:cognitive_mapping} presents cognitive mapping and its use to model business models.

\subsection{Software startups}
\label{ssec:software_startups}

The definition of software startup is not a consensus in the literature~\cite{Paternoster2014,Berg2018}, but the most common aspects are innovation and uncertainty~\cite{Berg2018}. Blank~\cite{Blank2007} proposed a definition generally adopted in practice:  a startup is an organization formed to search for a repeatable and scalable business model. This search for a viable business model for a novel software-intensive product is a defining aspect of software startups, contrasting these organizations to other development teams~\cite{Melegati2020b}, and being a significant contributor to the uncertainty these companies face~\cite{Giardino2014}. Therefore, in this paper, we use the following definition: software startups are organizations looking for a sustainable business model for an innovative product or service they develop where software is a core element.

However, these companies are not equal from an organizational perspective but, rather, display different development stages. Based in the literature, Klotins et al.~\cite{Klotins2019} proposed a life-cycle model to analyze the startups' progress composed of four stages: inception, stabilization, growth, and maturity. The first stage starts with the idea and ends with the first product release. In the next stage, the startup prepares to scale regarding technical and operational perspectives. In the growth stage, the startup aims to reach the desired market participation, and, finally, in the last stage, it progresses into an established company. In summary, during the early stages, i.e., inception and stabilization, teams focus on ``finding a relevant problem'' and ``a feasible solution.'' In the later stages, i.e., growth and maturity, the focus is on marketing and efficiency. We decided to initially focus on early-stage startups, given two reasons. First, the lack of testing assumptions about the customer and market represents a higher risk to the startup survival since, at this stage, the company generally does not have many resources. Second, in a later stage, the hypotheses obtained by the technique might have already been validated or refuted by the product usage in previous stages.

\subsection{Experimentation and Hypotheses Engineering}
\label{ssec:experimentation}

In software engineering research, experimentation has at least two meanings. Initially, it meant the use of scientific experiments to guide the research on software engineering. As an example, in a seminal book~\cite{Wohlin2012}, Wohlin et al., describe several empirical strategies to software engineering research. Recently, the term is used to describe the process of continuously validating product assumptions, transforming them into hypotheses, prioritizing, and applying the scientific method to test these hypotheses, supporting or refuting them~\cite{Lindgren2016}. This concept encompasses several techniques like iterations with prototypes, gradual rollouts, and controlled experiments~\cite{Fabijan2018} but also problem and solution interviews~\cite{Lindgren2016}. We adopted this second connotation in this paper.

In the literature, there are some models to describe experimentation in general, such as RIGHT~\cite{Fagerholm2017}, HYPEX\cite{Olsson2014}, and QCD~\cite{Olsson2015}. These models presented the process as cyclical approaches consisted of continuously executed steps~\cite{Melegati2019}: identify, specify, and prioritize hypotheses; design an experiment; execute it; analyze results; and update hypotheses accordingly. Nevertheless, these models do not describe how hypotheses could be systematically identified. 

In a previous paper~\cite{Melegati2019}, we called the step of identifying, specifying, and prioritizing hypotheses as Hypotheses Engineering. To explore currently available practices for hypotheses handling in the context of software startups, we performed a gray literature review~\cite{Melegati2021} on practitioners-produced documents. We identified 95 documents, analyzed them using thematic synthesis, and concluded that these practices could be divided into five activities: elicitation, prioritization, specification, analysis, and management.

\subsection{Available solutions}
\label{ssec:available_solutions}

Valuable pieces of prescriptive knowledge come from practitioner-oriented literature, such as the Customer Development~\cite{Blank2007} and the Lean Startup~\cite{Ries2011}. The latter had a huge success among practitioners and consisted of taking the founders' assumptions as hypotheses, building experiments to evaluate them, and based on the results, persevere or pivot, that is, change to another idea. One criticism against the Lean Startup is precisely its lack of operationalization. For instance, Bosch et al.~\cite{Bosch2013} proposed the Early-Stage Software Startup Development Model (ESSSDM) to tackle this problem. It consisted of three parts: idea generation, a prioritized backlog, and a ``funnel''. This funnel comprises four steps: problem validation, solution validation, small-scale minimum viable product validation, and large-scale minimum viable product validation. Through this funnel, ideas are validated. To generate ideas, the authors suggested exploratory interviews, brainstorming, or following potential customers to understand their needs.

In our exploration of the currently available practices to handle hypotheses in software startups~\cite{Melegati2021}, we identified several proposals to elicit hypotheses. We classified similar proposals into five categories. The most common category in the analyzed documents consisted of techniques based on canvases or maps, from which the best example is Assumption Mapping. This technique, recently proposed by Bland et al.~\cite{Bland2019}, consists of using a series of canvases, including the Business Model Canvas (BMC)~\cite{Osterwalder2009}, to create hypotheses. Although the BMC was initially based on an ontology systematically developed~\cite{Osterwalder2005}, Assumption Mapping has been neither derived nor evaluated scientifically.

The second most common category was the use of a pre-defined set of questions or aspects to consider. For instance, \textit{which customer problems are to be solved? Can our product solve the customer problem}. Then, in the following category, proposed techniques suggested the execution of team sessions where members could reach hypotheses based on facilitation techniques. In the fourth category, practices proposed describe individual techniques the founder could use to reach hypotheses such as ``the five whys''. Finally, another suggestion is to use problem or solution interviews to elicit hypotheses. Although there are several available practices for hypotheses elicitation, none of them were systematically derived nor evaluated.	

\subsection{Cognitive mapping}
\label{ssec:cognitive_mapping}

The business model concept has a plethora of different definitions used in academic literature~\cite{Zott2011}. Furnari~\cite{Furnari2015} described two theoretical perspectives in business model research: an activity-based perspective that describes a business model as ``a system of activities that firms use to create and capture value,'' and a cognitive perspective considering it as a cognitive instrument to represent those activities. 

Based on the cognitive perspective, Furnari~\cite{Furnari2015} proposed the use of cognitive maps to represent business models. Cognitive maps are visual representations of causal aspects of a person's belief system as a graph where nodes represent the concepts individuals use and arrows, causal links between them~\cite{Furnari2015}. The arrows are generally labeled according to the type of relationship: `+' for a positive one, `-' for a negative one, and `/o/' for a neutral one. Cognitive maps are supported by Kelly's Personal Construct Theory~\cite{Eden1988} which states that a person looks at the world through patterns or templates, called constructs, that this person creates and, in which, tries to fit the reality~\cite{Kelly2002}. Kelly also describes the person-as-a-scientist idea: ``as a scientist, man seeks to predict, and thus control, the course of events'' and these constructs ``are intended to aid him in his predictive efforts''~\cite{Kelly2002}. Brannback et al.~\cite{Brannback2009} have already discussed the relationship between cognitive mapping and the personal construct theory with entrepreneurship. The authors argued that ``an entrepreneur needs to make sense of his/her reality to predict and control - to find and solve problems''~\cite{Brannback2009}.

One essential aspect of software startups is the founders' influence on the product definition. Seppanen et al.~\cite{Seppanen2016} investigated the competencies of initial teams in software startups. They observed a strong influence from founders on the actions and competencies related to the business and product creation in these nascent companies. 

Based on what we have described so far, software startups' business models are strongly influenced by how their founders perceive, mentally model the environment and how they use these models to predict the market and how the future product will behave. Research has shown that this influence is strong enough to prevent the use of experimentation. For instance, while investigating the limited adoption and implementation of experimentation in software startups, Melegati et al.~\cite{Melegati2020c} identified as an inhibitor the fact that founders are often overconfident with the idea, deeming experiments to evaluate it as unnecessary and focusing on developing the solution. Giardino et al.~\cite{Giardino2014} argued that this focus on launching the product is one of the key challenges early-stage software startups face.

Cognitive mapping could be used to materialize the assumptions regarding several aspects of the software product, such as customers, market, and technology, on which startup founders base their ideas and put them in a position to be challenged. As Eden~\cite{Eden1988} points out: ``by seeing their own ideas in this form of visualization [people] are being encouraged to `change their mind.''' This technique has been used in Software Engineering, for instance, in problem structuring in requirements engineering~\cite{Rooksby2006} or a decision model for distributed software development with agile~\cite{Almeida2011}.

The methods to elicit cognitive maps can be divided into two groups depending on how data is obtained~\cite{Hodgkinson2004}. Initially, researchers used documents or other sources of evidence to perform content analysis to develop these maps. Another way is through direct methods where researchers develop the map \textit{in situ} by interacting with subjects from whom the cognitive map is elicited. These direct methods can employ two divergent approaches: pairwise judgments of causal relationships or capture through visual forms. In the first form, subjects answer questionnaires where all combinations of concepts are evaluated, and based on the answers, a map is built. In the second form, with the subject's help, a facilitator builds a visual representation using paper and pencil or software solutions. The pairwise approach has better coverage at the expense of being more difficult, less engaging, and less representative than the freehand technique~\cite{Hodgkinson2004}. Since the startup context is defined by time and resource constraints~\cite{Berg2018}, an effective approach targeted to these companies should not be time-consuming. Therefore, a visual approach is more suitable.

\section{Research method}
\label{sec:research_method}

Given our research goals and question, we aim to solve a real-world problem. Instead of trying to understand how a defined phenomenon unfolds, our goal is to develop an artifact to act on the world. In this regard, Design Science Research (DSR) is a suitable method. This approach is often used in Information Systems research as shown by the several methodological guidelines (e.g., Hevner et al.~\cite{Hevner2004}, Peffers et al.~\cite{Peffers2007}, and Wieringa~\cite{Wieringa2009}) and even a special issue in \textit{MIS Quarterly}~\cite{March2008}. Although its use is often not explicitly mentioned in Software Engineering research, in an analysis of awarded papers in the \textit{International Conference on Software Engineering}, Engstrom et al.~\cite{Engstrom2020} showed that most of these studies actually could be classified as DSR although not explicitly using the term. More recently, though, researchers have explicitly used the methodology to tackle problems in software engineering like gamification (e.g.,~\cite{Morschheuser2018}) and requirements (e.g.,~\cite{Benfell2020}).

In this research, we followed the guidelines proposed by Hevner et al.~\cite{Hevner2004,Hevner2013}.  According to the authors, DSR seeks to develop innovative artifacts relying on existing kernel theories ``that are applied, tested, modified, and extended through the experience, creativity, intuition, and problem solving capabilities of the researcher''~\cite{Hevner2004}. These artifacts could be constructs, models, methods, or instantiations. Constructs represent the language used to describe the world, and models use them to represent real-world situations. Methods define processes to guide how to solve problems and, finally, instantiations demonstrate how the previous elements could be used in a scenario. Based on this classification, the artifact developed in this study is a method allowing early-stage software startups to identify hypotheses to guide experiments.

\begin{figure}
		\centering
		\includegraphics[width=\columnwidth]{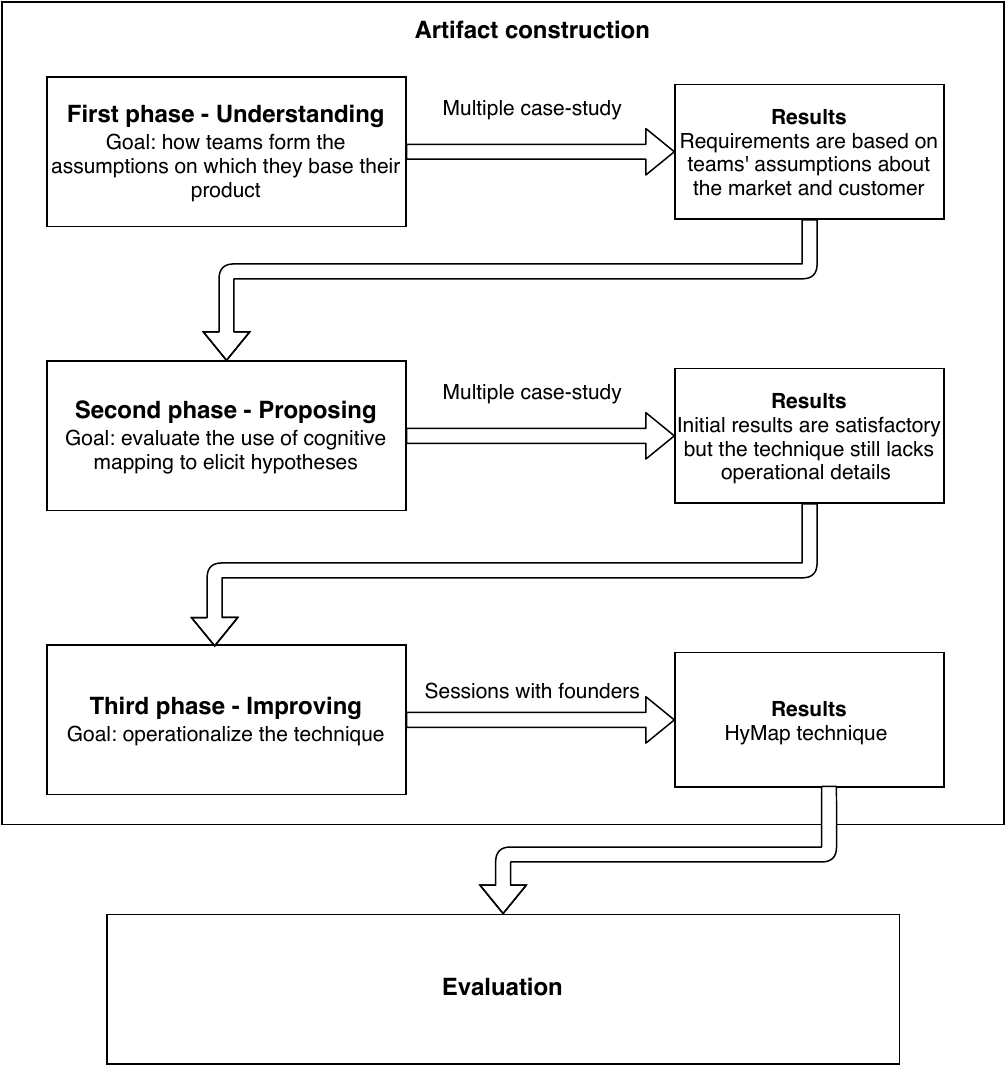}
		\caption{The research method including the artifact construction process and its evaluation.}
		\label{fig:research_method}
	\end{figure}

In a DSR study, it is essential to describe the artifact development process and its evaluation~\cite{Hevner2004}. For this study, the design artifact development process consisted of three phases where the first two were already described in a previous paper~\cite{Melegati2020}. Fig.~\ref{fig:research_method} summarizes the research method.
	
The first phase, Understanding, had the goal of understanding how teams form the assumptions on which they base their products. To achieve this goal, we performed a multiple-case study with two early-stage software startups. Our results indicate that requirements are based on the team's, especially the founder's, assumptions about the market, and customer behavior. In the second phase, Proposing, we used cognitive mapping to make founders' assumptions explicit. At this stage, the technique consisted of using boxes and the arrows as described in cognitive mapping and the employment of an open-ended talk where the founder described her understanding, and a facilitator drew the map. We assessed this first version of the method in two other software startups. Our results indicate that this approach could base a comprehensive practice to elicit hypotheses in software startups. However, it still lacked some operationalization. In the third phase, Improving, we improved the technique by defining specific notations for different concepts (customer, product, and features) and creating a list of questions to guide the cognitive map creation.

Since a startup is a complex phenomenon with many variables, such as founders' background, product, and market, and a blurred boundary between phenomenon and context exists, a case study is a suitable choice to evaluate a technique for these companies. According to Yin~\cite{Yin2003}, one rationale for this research approach is the representative or typical case. Therefore, we selected software startups where, as mentioned before, the founder is the one who had the initial idea. Besides that, we followed Klotins et al.'s life-cycle model~\cite{Klotins2019} and selected startups in the inception and stabilization phase.

We guided the development and evaluation processes according to defined criteria: utility, quality, and effectiveness. First, to fulfill the utility criteria, the artifact aims to have ``value outside the development environment''~\cite{Hevner2013}. Therefore, using the artifact, founders should be able to create hypotheses for real situations, that is, for startups other than those that participated in the study. Utility is associated with the perceived usefulness, which is generally used to explain the adoption of software development methodologies (e.g.,~\cite{Riemenschneider2002,Hardgrave2003}) and technology in general~\cite{Davis1989}. Perceived usefulness ``refers to the degree to which a developer expects that following a methodology will improve his or her individual job performance''~\cite{Hardgrave2003}. In the context of experimentation in software startups, we can operationalize this concept by obtaining hypotheses to build experiments.

Regarding quality, artifacts can be evaluated from several attributes, such as completeness, usability, and consistency~\cite{Hevner2004}. In this study, we selected three attributes to focus on this evaluation: ease of use, clarity, and self-containedness. Since our ultimate goal is to impact real startups, we should consider this method's future adoption. In this regard, taking the artifact as innovation, complexity is a factor influencing adoption~\cite{Rogers2010}. Therefore, founders must perceive the technique as easy to use, and its concepts and the process steps as clear. It should be self-contained, that is, it should work independently to who is applying or facilitating it, providing all the details to proper use, allowing anyone to use it rather than depending on the facilitator's experience. Finally, the method description should be clear, making its comprehension straightforward.

Finally, the artifact should be effective; that is, producing the expected result. In our case, the goals are to reveal hidden assumptions, that founders had about the product's environment and potential customers, and to systemically elicit hypotheses that would be the basis for experiments.

Following this method, we aim to satisfy the seven guidelines for proposed by Hevner et al.~\cite{Hevner2004} for DSR:

\begin{enumerate}
	\renewcommand{\labelenumi}{G\arabic{enumi}.}
	\item \label{guide:1} Design as an artifact: DSR must produce a viable artifact;
	\item \label{guide:2} Problem relevance: the goal should be to develop a solution to relevant problems;
	\item \label{guide:3}  Design evaluation: the ``utility, quality, and efficacy'' of the artifact should be rigorously demonstrated;
	\item \label{guide:4}  Research contributions: the DSR project should provide ``clear and verifiable contributions'' regarding the ``design artifact, design foundations, and/or design methodologies'';
	\item \label{guide:5}  Research rigor: DSR should apply rigorous methods both in the construction and in the evaluation of the artifact;
	\item \label{guide:6}  Design as a search process: the DSR process is inherently iterative and the search for the best, optimal solution is unfeasible. The goal should be feasible, good designs representing satisfactory solutions.
	\item \label{guide:7}  Communication of research: DSR solutions should be presented effectively.
\end{enumerate}

Since the target artifact is a method, we satisfy~\prettyref{guide:1}. Our argument about the importance of experimentation to early-stage software startups fulfills~\prettyref{guide:2}. In the following sections, we describe the design artifact (\prettyref{guide:5}), its development process (\prettyref{guide:6}), and evaluation (\prettyref{guide:3}). In Section~\ref{sec:discussion}, we discuss the research contributions (\prettyref{guide:4}). This paper and previous one~\cite{Melegati2020} communicate our results (\prettyref{guide:7}).

\section{Artifact design process}
\label{sec:development_process}

This section describes the three phases of the artifact construction process, Understanding, Proposing, and Improving, in detail, including the research method, data collection, analysis, and results obtained.

\subsection{First phase - Understanding}
\label{ssec:cycle_0}

This phase's goal was to understand how teams form the assumptions on which they base their products. We selected two startups called from now on as A and B, located in an Italian technological park. At the moment of data collection, startup B participated in the incubation process, while startup A only used the space available.

Data collection consisted of semi-structured interviews that followed a pre-defined guide. For both cases, we interviewed the founders and, for case B, also the software developer. The interview questions aimed to understand the interviewees' background, the idea, the motivation to build the product, and how they changed throughout the company history. Since the goal was to understand from where the assumptions used to create the product came, data analysis consisted of explanation building where cause-effect relationships are sought~\cite{Runeson2012} looking for an explanation for the cases~\cite{Yin2003}.

\subsubsection{Case A}

At the time of data collection, the startup was developing a library to be added to software development projects. The company planned to provide a dashboard to show live software run-time issues, like exceptions, detected or inferred from data collected within the target system. The dashboard would also show solutions to similar problems found on websites focused on programming issues, such as Stack Overflow, and a list of freelance developers that could solve the problem. In some cases, the system would be able to fix some issues automatically. The startup team consisted of five people working part-time on the project spread across software development, business plan, and marketing strategy. 

The founder had worked as a software development consultant for an extended period. While participating in third-party projects, he observed that such a tool could help him work more effectively. Besides that, he believed that the technical level of software developers was decreasing. Therefore, it would make sense to develop such a tool.

At the time of data collection, the startup had an initial prototype consisted of a dashboard with some dummy data and a website presenting the idea. 

\subsubsection{Case B}

The startup was running a website to help hotel owners and managers to find the best software solutions for their businesses. Its initial focus was on the Italian market, but it aimed to reach international markets. The team consisted of two founders/partners, one developer who founded the company but left the partnership, and an intern to help with administrative tasks. We performed interviews with the founder who had the original idea and the developer.

The interviewed founder had a background in online marketing. He had worked in a company that handled web marketing and websites before staying twelve years in a large web agency. In his last job, he worked as the director of the company's technology business unit. Throughout his work life, he had extensive contact with the tourism sector, especially the hospitality industry. 

The founder told having the idea based on the needs he observed from hotel owners, which have many technological tools available to run the business, and software vendors, that have to reach these customers. The founder's inspiration was American software review websites that compare different software solutions to particular market segments and the lack of a specific one for the hospitality sector. Therefore, the original idea was to list available software solutions with users' reviews, attract hotel owners to the website, and receive a commission from vendors for each interested customer that visited their websites.

The founder said that, after the original version had gone online, the team started observing the website usage data and realized it was not going as expected. The team observed that the hotel owners were not able to compare different software solutions because these products rarely have the same set of features, and, sometimes, hotels needed more than one software system to fulfill their requirements. Then, the startup changed the website to ask the hotel owner to fill an online form giving details about the business. Based on the information provided, the system would match the solutions adapted to the business needs. 

In the interview with the developer, it was clear that his influence on the idea was limited. He thought that it was a good idea but did not have experience in the market. He trusted the founders regarding the business and focused on developing the solution. 

At the moment of data collection, the startup had its customer base growing, and it was close to break-even. It was looking to expand to other markets.

\subsubsection{Cross-case analysis}

\begin{figure}
	\centering
	\includegraphics[width=0.6\columnwidth]{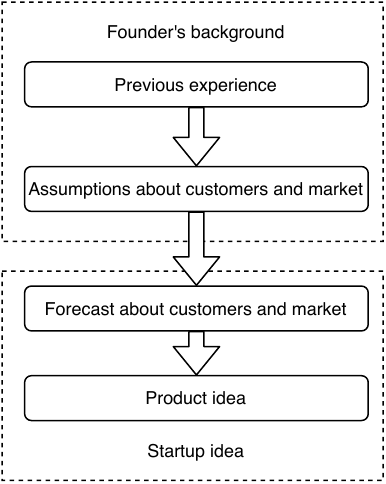}
	\caption{The process of idea creation. The dashed lines represent the previous understanding that the background of the founder led to the idea. Adapted from~\cite{Melegati2020}.}
	\label{fig:idea_creation}
\end{figure}

A common aspect between cases A and B is that, based on their previous experience, the founders formed a set of beliefs to explain how a specific market functions and its customers behave. Based on this understanding, founders predict how potential customers would behave when exposed to their product or service. For instance, in startup B, the founder considered that hotel owners wanted to buy software solutions and that they were able to compare the different alternatives and select the best suited for their cases. Based on that, the founder foresaw that a website with a list of available software solutions would be useful to hotel owners. They would be able to see all the solutions and select the one that would fit their needs. Fig.~\ref{fig:idea_creation} summarizes this process: the founder's previous experience leads to assumptions about the customers and the market, used to forecast the behavior of customers , and a new product idea. This process is related to the founder being the innovation owner and her experience being the motivator for the startup product idea, as discussed in the literature~\cite{Seppanen2016}.

In startup B, after the software was ready and the website went online, usage data showed that the results were not as predicted. Hence, the founder had to update his assumptions about the customers and change the product accordingly. This new ``implicit theory'' has emerged from the experiment result and led the company to better results within the market. Nevertheless, to reach this stage, the startup spent resources developing the whole product that could have done earlier if the team had analyzed the customers.

\begin{figure}
	\centering
	\includegraphics[width=\columnwidth]{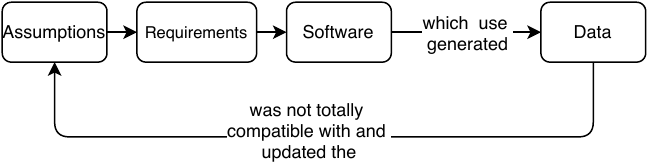}
	\caption{The founder's assumptions being updated as shown in~\cite{Melegati2020}.}
	\label{fig:feedback}
\end{figure}

Such rearrangement exposed an implicit process model for development in software startups. In such a process, the founder's assumptions guide the elicitation of requirements, and the data generated by the software usage may impose changes on this set of assumptions. Then, the founder uses such an updated representation of the world to elicit new requirements. Fig.~\ref{fig:feedback} depicts this process through a causal chain, ``a research-constructed linear display of actions and/or states that suggests a plausible, interrelated sequence of events''~\cite{Miles2014}.

\subsection{Second phase - Proposing}
\label{sssec:cycle_1}

Based on the first phase results, a hypothesis elicitation approach should make explicit the founders' underlying assumptions about the customers and the market. In this phase, our goal was to evaluate the feasibility of using cognitive mapping to make these assumptions explicit. To do so, we adapted the approach proposed by Furnari~\cite{Furnari2015}. Using a whiteboard to depict the current status of the mapping and, with the founder's help, we aimed to create a cognitive map representing how and why the founder believes the startup's business model works. The detailed steps were:

\begin{enumerate}
	\item ask the founder to describe the business model concerning the value proposition and customers;
	\item extract concepts and causal relationships;  
	\item inspect each concept to see if they were, in reality, not based on an underlying assumption. 
	\item check with the founder if the map represented the way the problem was understood at the moment. 
\end{enumerate} 

We evaluated this initial proposal in two other Italian software startups, C and D. We performed interview sessions following the defined protocol below:

\begin{enumerate}
	\item Present the concept of hypotheses and how Lean Startup is related to it.
	\item Ask the interviewee to describe his business or product idea, especially regarding customer segments and value proposition.
	\item Ask on which hypotheses the founder believed his idea is based.
	\item Using a whiteboard and interacting with the founder, draw a cognitive map until the map represented the understanding of the market.
	\item Create a list of hypotheses based on cognitive mapping and compare it with the initially created list.
	\item Ask feedback on the process to the founder.
\end{enumerate}

Below, we describe the results for each case.

\subsubsection{Case C} 

Case C was an early-stage software startup planning to develop a digital mentor for software developers to increase their happiness and satisfaction. The product would adapt itself to each developer's needs. Companies interested in improving their developers' productiveness would pay a fee to make the solution available to their teams. 

When asked about hypotheses, the founder mentioned some they had already worked with and others they were planning to evaluate. The first hypothesis was that software development teams could not organize themselves. Through some interviews, it got invalidated, and they changed from the initial idea to the current one. The following hypothesis or, how the interviewee called, \textit{``exploration''} was to understand if software developers care about soft skills. When asked about other hypotheses, the founder said that she was waiting for another round of tests.

\begin{figure}
	\centering
	\includegraphics[width=.8\columnwidth]{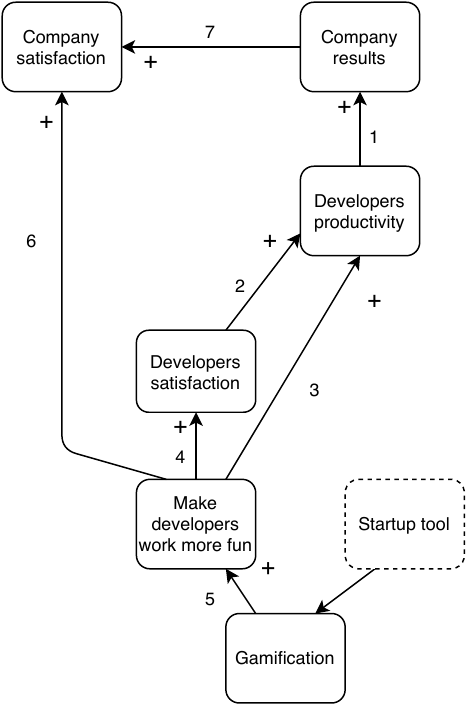}
	\caption{Cognitive map created during interview with the founder of startup C. Numbers were included here to identify the hypotheses.}
	\label{fig:case_c_cognitive_map}
\end{figure}

Fig.~\ref{fig:case_c_cognitive_map} displays a representation of the cognitive map derived for this case. Through this process, the founder stated that the main element to increase developers' productivity would be making their work more fun through gamification.

The arrows in the figure imply seven hypotheses: 1) developers productivity improves the company results; 2) developers satisfaction improves developers productivity; 3) making the development work more fun improves the developers' productivity and 4) the developers satisfaction; 5) gamification could make developers' work more fun; 6) making the development work more fun would increase the company satisfaction; 7) increasing the company results will improve its satisfaction. 

Although some identified hypotheses are straightforward and may not demand a proper experiment to be considered validated, the founder acknowledged that the \textit{``[they] have to see if the correlation between having fun and the productivity [exists], that is a major risk.''}

\subsubsection{Case D}

Case D was developing a software solution to improve the perceived quality of connection to the Internet, especially for locations where such quality level is low, such as small mountain villages. Through an innovative approach, the solution would make the network status transparent, enabling the user to adapt it to their needs and improve the quality of service. At the request of the interviewee, details of this product are not included in this article.

\begin{figure}
	\centering
	\includegraphics[width=\columnwidth]{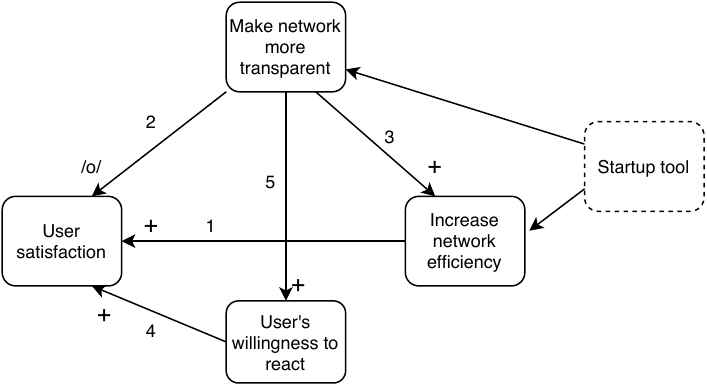}
	\caption{Cognitive map created during interview with the founder of startup D. Numbers were included here to identify the hypotheses.}
	\label{fig:case_d_cognitive_map}
\end{figure}

At the beginning of the interview, the founder explained that he considered that the main hypothesis concerned with the size of the low-quality area and if Internet providers were willing to quickly fix the problem. He mentioned talking to many potential customers regarding the solution, and most of them would like to have the solution. Fig.~\ref{fig:case_d_cognitive_map} depicts the cognitive map developed in the session. 

From the arrows, there are five implied hypotheses: 1) increasing the network efficiency will improve user satisfaction, 2) making the network more transparent will not decrease user satisfaction, 3) making the network more transparent will increase the user's willingness to react [to the bad quality of Internet connection], 4) the users' willingness and ability to react will increase user satisfaction, and 5) making the network more transparent will increase the user's willingness to react.

When confronted with the hypotheses, the founder mentioned they had already thought about them before. Nevertheless, using his words, the process \textit{``made them explicit and more structured.''}

Although the results were promising, we observed that the process of cognitive map elicitation was not repeatable because it was highly dependent on the interviewer's experience. Besides that, the interviewer felt the lack of guidance on properly conducting this step. Another aspect that we observed was the lack of uniformity in the boxes' content: some were concepts or nouns, but others were actions, events. Based on that, we improved the technique in the next  phase. Regarding the expected attributes defined in Section~\ref{sec:research_method}, the artifact was useful, easy to use, and effective, but we should improve its qualities: self-containedness and clarity. 

\subsection{Third phase - Improving}
\label{sssec:cycle_2}

Based on the previous phase results, the need for a more systematic approach was evident. To achieve this goal, we focused on developing a proper visual language to build the maps and a systematic method to elicit them from founders. Since the artifact development is a design search process~\cite{Hevner2013}, we performed a series of map elicitation sessions with potential founders. The subjects in this phase not necessarily created a startup but had an idea that, in their opinion, could potentially be the basis of a new solution. After each session, we evaluated the process and improved the artifacts to make the language and process more precise. The sessions occurred online, and the researcher shared his screen where he drew the cognitive map using the software Diagrams.net\footnote{https://app.diagrams.net/} with the interviewee's help. Once the results reached a satisfactory level, we considered the artifact design process completed.

Following this approach, we performed three sessions. While the first two were Brazilian entrepreneurs, the third was based in Italy. In each session, we tried to improve the visual language and the map elicitation process based on the previous session results. Table~\ref{tab:cycle2} describes the improvements we applied to the technique in each session and the respective results.

\begin{table*}[!ht]
	\renewcommand{\arraystretch}{1.5}
	\caption{Sessions performed on the third phase of artifact creation.}
	\label{tab:cycle2}
	\begin{tabular}{cp{.3\textwidth}p{.3\textwidth}p{.3\textwidth}}
		\hline
		Session & Visual language improvements & Process improvements & Results \\ \hline
		1 & Using a circle to differentiate the customers' problems from other concepts on the map. & An initial set of questions to guide the map elicitation (product name and customers' problems the solution aimed to solve) and the idea of employing an iterative approach to connect these two elements. & Much clear guidance for the interviewer with respect to the previous phase. But the process of asking the interviewee about the customers and their problems was still not straightforward. \\ \hline
		2 & Splitting the customers and their problems by using circles to represent customers or users and boxes to represent their problems, similar to other concepts in the map. & Questions changed accordingly to changes in the language: specific questions were added to ask the customer segments and their problems. & Improved the process but, based on the analysis of the elicited maps, using the same shape to represent software features and value concepts was confusing. \\ \hline
		3 & Dashed-line boxes to represent features, differentiating them from other concepts. & None. & Satisfactory results. \\ \hline
	\end{tabular}
\end{table*}

\section{HyMap: Hypotheses Elicitation using Cognitive Maps for Early-Stage Software Startups }
\label{sec:artifact_description}

HyMap, the method developed in this study, is a process, conducted by a facilitator, to draw a cognitive map and extract hypotheses from it based on a visual language to depict the map. 

\subsection{A visual language to represent startup founder cognitive maps}

To depict the cognitive map that supports the elicitation process, we developed a visual language adapting the cognitive mapping symbols described in Section~\ref{ssec:cognitive_mapping} to differentiate the concepts represented. It consisted of the following elements:

\begin{itemize}
	\item Circles represent customer segments.
	\item An ellipsis is used to denote the proposed solution.
	\item Dotted-line boxes portray the software features.
	\item Solid-line boxes represent concepts, either physical or abstract, and are filled with nouns.
	\item Arrows connect boxes or circles to boxes and represent relationships among them. There are three types of relationships: offering, influence, and perception, that are defined by the types of the shapes connected. Offering arrows connect the solution and its features to represent that the product offers those functionalities. Influence arrows are similar to those in conventional cognitive maps, as mentioned  in Section~\ref{ssec:cognitive_mapping}, and represent that one concept influence the intensity of the other. They should be labeled with one sign: `+', `- ', `/o/' to denote its type: respectively, if one concept increases the other, decreases, or does not influence. Perception arrows are those that connect the customers with their problems and represent that customers perceive those concepts as issues.
\end{itemize}

\begin{figure}
	\centering
	\includegraphics[width=\columnwidth]{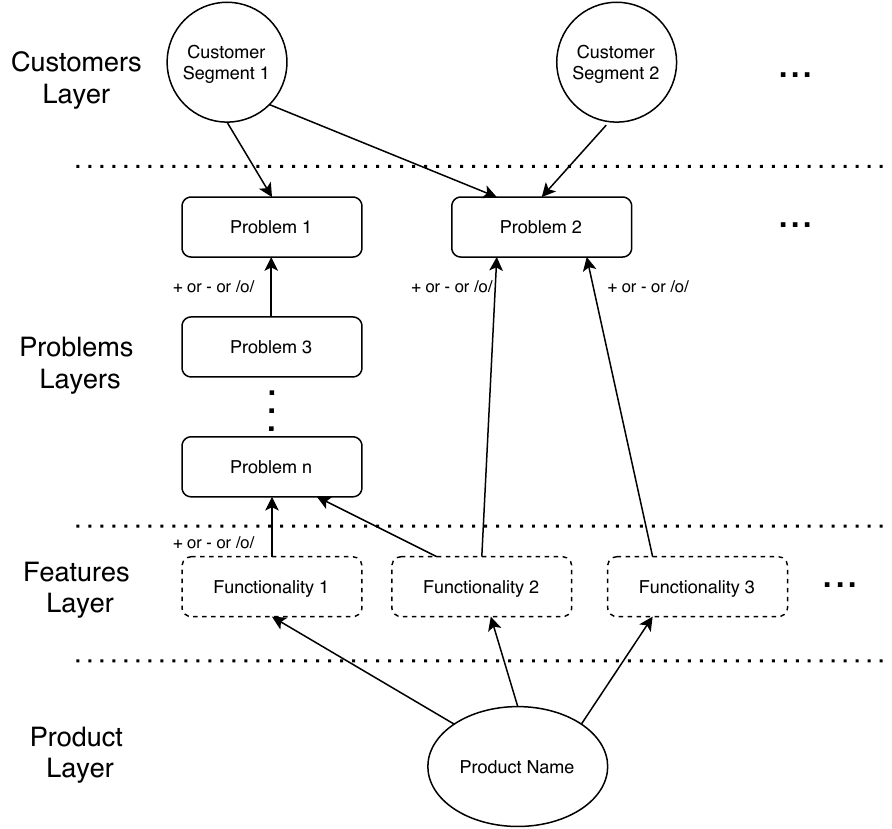}
	\caption{A template for the HyMap map.}
	\label{fig:hymap}
\end{figure}

There are no restrictions on the number of inbound or outbound arrows from boxes, but they must represent an acyclical graph. Such a pattern for the construction leads to layers of elements in the map, as shown in Fig.~\ref{fig:hymap}. In the Product layer, we represent the product. In the Features layer, we represent the features the founders expect for the product. In the Problems layers, one or more layers of elements represent the aspects founders think the product features will solve. Finally, in the Customer layer, we represent the expected customers and users for the product.

\subsection{Eliciting startup founder cognitive maps and hypotheses}
\label{ssec:elicitation_process}

The first step to reach hypotheses is to elicit the founder's cognitive map. To reach this goal, we propose an iterative approach where, based on the map's current status, the founder should analyze each of the relationships (arrows) and consider if there are underlying concepts. A facilitator helps the founder in this process by following a pre-defined set of questions, drawing the map according to the visual language, and clarifying the questions if the founder does not understand them. This way, the founder focuses on the business questions while the facilitator that had previously used the technique asks the founder questions and handles the drawing. The facilitator could also ask for clarifications when details are not clear. At the beginning of this process, an initial map should be created based on the following questions:

\begin{enumerate}
	\item What is the product/solution name?
	\item What are the customers targeted by the solution?
	\item For each customer, what are the aspects the actor expects to improve using the solution?
	\item Which are the solution features envisaged, and which aspects in the previous step do they fulfill?
\end{enumerate}

Based on these questions' answers, the facilitator put the respective shapes on the map and connect them with arrows. Then, for each arrow, the founder should judge if there are concepts implicitly used to explain that relationship. Some questions useful in this step are \textit{how?} and \textit{why?}. If a new concept emerges, a new box representing it is added along with new relationships (arrows). This process is repeated iteratively until the founder is comfortable that no new concepts should be added. A useful question to evaluate if this process is saturated is if it is possible to create a simple experiment to evaluate that relationship. Additionally, the founder must evaluate if the new concepts added are related to other concepts already present on the map. Throughout the process, the founder should constantly assess if the map being created is coherent to her understanding of the customer and market. To refine the map, the founder can add, remove, or substitute elements.

Once the cognitive map is finished, we can say that each relationship represents an assumption the founder has about its targeted customer, value proposition, and product. Based on them, she can formulate hypotheses based on which she can create experiments. These experiments can be pieces of software but also interviews, questionnaires, or other techniques. 

Arrows originating in different layers, as shown in Fig.~\ref{fig:hymap}, represent diverse types of hypotheses that demand different templates while crafting the hypotheses. Although the definition of systematic templates for hypotheses is beyond this paper's scope, we defined a simple template for each type. Of course, the templates are only guidelines, and a final inspection of the wording is necessary to create well-formed phrases. Below, we describe each type and the corresponding template. 

\begin{enumerate}

\item Arrows from the product to the features layer generate hypotheses regarding the team's capability to develop that feature, the so-called product hypotheses. A simple template for this type is: ``the team developing $<$product name$>$ is capable of implementing $<$functionality$>$''.  

\item Arrows starting from the features layer to the problem layers, or those restricted to the problem layers, represent value hypotheses. In this case, a suitable template is ``$<$Functionality or problem$>$ $<$increases, decreases or does not affect$>$ $<$problem$>$''. 

\item Arrows connecting customers to the problem layers lead to problem hypotheses, that is, if that problem is a real ``pain'' for the customers, that will make her pay for the solution. An initial template for this type of hypothesis is ``$<$Customer segment$>$ $<$has/would like to$>$ $<$problem$>$''.

\end{enumerate}

\section{Evaluation}
\label{sec:evalution}

In this section, we describe the HyMap evaluation, including the method employed and the results obtained.

\subsection{Method}

To evaluate HyMap, we executed a protocol similar to the one employed in the second phase of the artifact construction but online as in the third phase. To evaluate the artifact's ease of use and self-containedness, a different researcher, with no previous experience with startups, from the one that performed the sessions on the construction phase, was responsible for facilitating the sessions by following the protocol and asking the pre-defined questions. The other researcher acted as an observer during the elicitation sessions. Another difference was that we divided the protocol into two steps: first, the facilitator, with the help of the founder, created the map; then, we created the hypotheses list offline strictly from the maps, where each arrow represented a hypothesis, written using the template described in Section~\ref{ssec:elicitation_process} and sent them to the founder. In a second session, we performed an interview to get feedback on the hypotheses and the process. If the founder was not available for a second interview, we sent a questionnaire.  We also used this instrument as a guide if we performed the second interview. In both situations, we started asking for each hypothesis if the founder believed the hypothesis had been validated, that is if the founders have evidence to support that hypothesis. For instance, if the founder had talked to customers and concluded that the hypothesis valid. In case the answer was positive, we asked how the founder reached that conclusion. Then, we asked how the founder perceived the risk to the business if that hypothesis was not valid. Then, we asked feedback about the process through four open questions:

\begin{enumerate}
	\item Do you consider the process useful? Why?
	\item Do you consider the process/the diagram easy to implement/use? Why?
	\item Do you consider the process and the diagram clear?
	\item Did the process lead you to think about something you had not considered before but will pay more attention for in the future of your startup?
\end{enumerate}

To sample the cases, we employed a theoretical approach based on the different startup stages, as described in Section~\ref{sec:literature_review}. Since the focus of HyMap is on early-stage startups, we aimed at companies in the inception and stabilization stages. For the inception stage, we selected two cases, one at the beginning and the other at the end of the stage given that, at the end, the startup has an initial version of the product. Hence, with one case in the stabilization stage, we selected three cases. We followed a convenient approach, using our contacts network, to recruit interested startups. 

\subsection{Results}

We performed the planned case study in three startups that we referred to as E, F, and G, ordered by the development stage they are at the moment of data collection: beginning and end of inception and stabilization stages, respectively. Below, we describe the companies and the results we obtained for each in detail. To preserve the startups' privacy, a request that founders often make, we do not explicitly put the product or startup's name in the descriptions or the cognitive maps.

\subsubsection{Case E}

\begin{figure*}
	\centering
	\includegraphics[width=.8\textwidth]{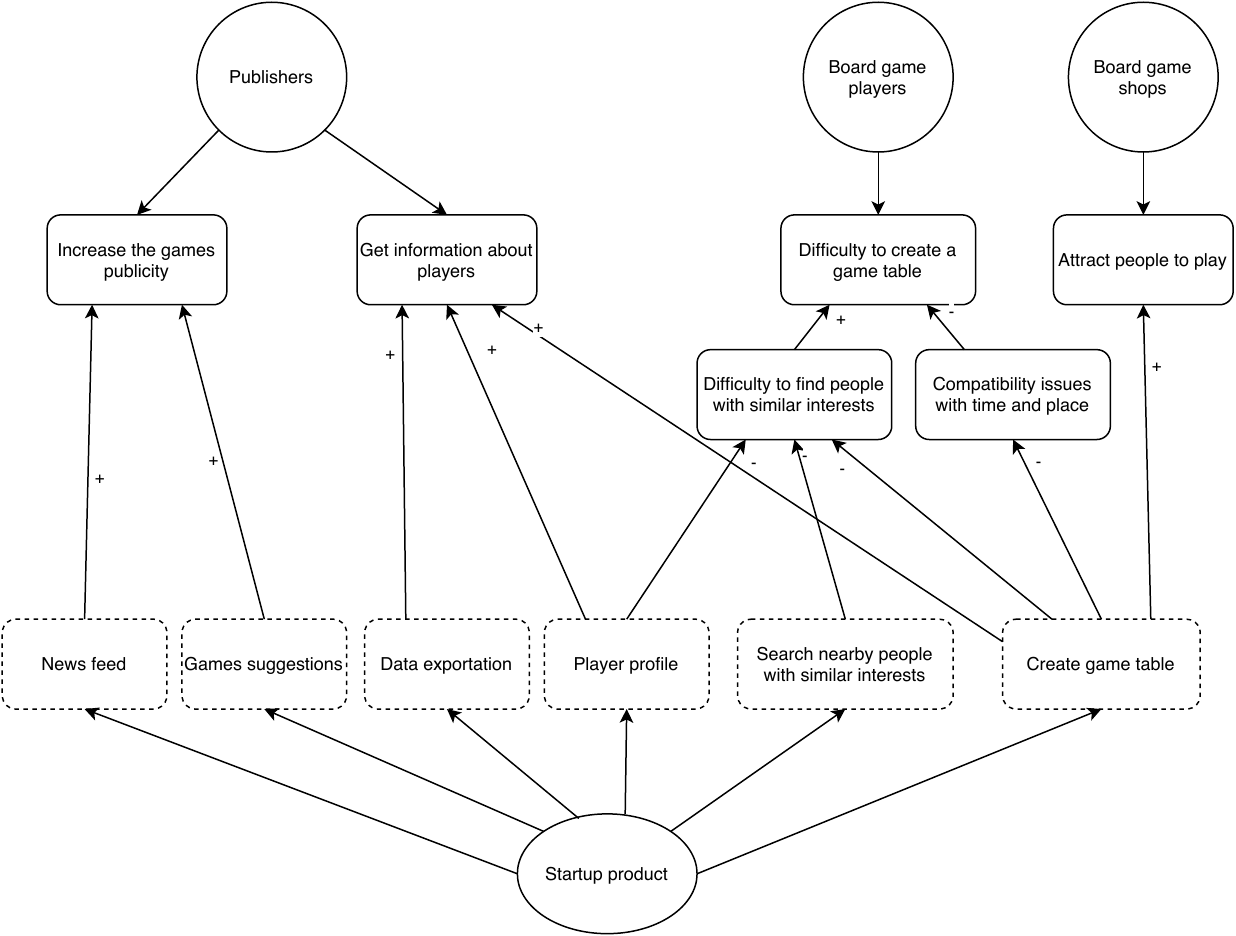}
	\caption{Cognitive map created for startup E.}
	\label{fig:case_e_cognitive_map}
\end{figure*}

Case E was a Brazilian startup planning an app to connect board game enthusiasts to meet and form groups to playing sessions. The startup also planned to provide services to board game shops for attracting people to play on their premises and publishers for promoting their games. At the time of data collection, the startup had already created the brand and started building an online presence. Because of the 2020 coronavirus pandemic, the development halted, and the startup lost all team members but the founder. Therefore, we classify the startup at the beginning of the inception stage. The interview performed with the founder resulted in the cognitive map depicted in Fig.~\ref{fig:case_e_cognitive_map}. 

Based on the cognitive map, we identified 22 hypotheses of which four were related to problems (e.g., ``board game players have difficult to form game tables''), 12 to value (e.g., ``the search of nearby people with similar interests decreases the difficulty to find people with similar interests''), and six to the product (e.g., ``the development team is capable of implementing the search of nearby people with similar interests'').

Regarding the problem hypotheses, the founder said that one (``board game players have difficult forming game tables'') represented a high risk to the business if it was not valid. The founder claimed to have validated them through her own experience within the field, and with offline and online surveys. The other three problem hypotheses represented a medium risk, and she claimed to have validated them through talking to shops and publishers. Out of the 12 value hypotheses, the founder considered eight with high risk to the business and four with medium risk. She believed that all value hypotheses were validated except one: ``creating game tables through the app facilitates bringing people to play at the shop.'' For the 11 value hypotheses the founder considered validated,  we grouped similar strategies she claimed to have employed to validate these hypotheses. Since she mentioned more than one strategy per hypothesis, the sum of occurrences is larger than the total of hypotheses. She mentioned that validation came with offline and online surveys for six of them, four from the comparison with similar tools, three from resembling business models, and three from her own experience with the market. Finally, regarding product hypotheses, the founder considered all with high risk except the one regarding news feed (low risk) and suggestions (medium risk). However, they were not validated so far because of the lack of a development team.

Regarding the answers to the feedback questions, the founder considered the process useful because ``it helped me to visualize other possibilities of the project and also to view points I had not observed before.'' She considered the diagram and process easy to use because they ``follow a practical and objective logic.'' She also deemed the diagram and process clear, and they made her think on other points, although she did not clarify which points.

\subsubsection{Case F}

\begin{figure*}
	\centering
	\includegraphics[width=.8\textwidth]{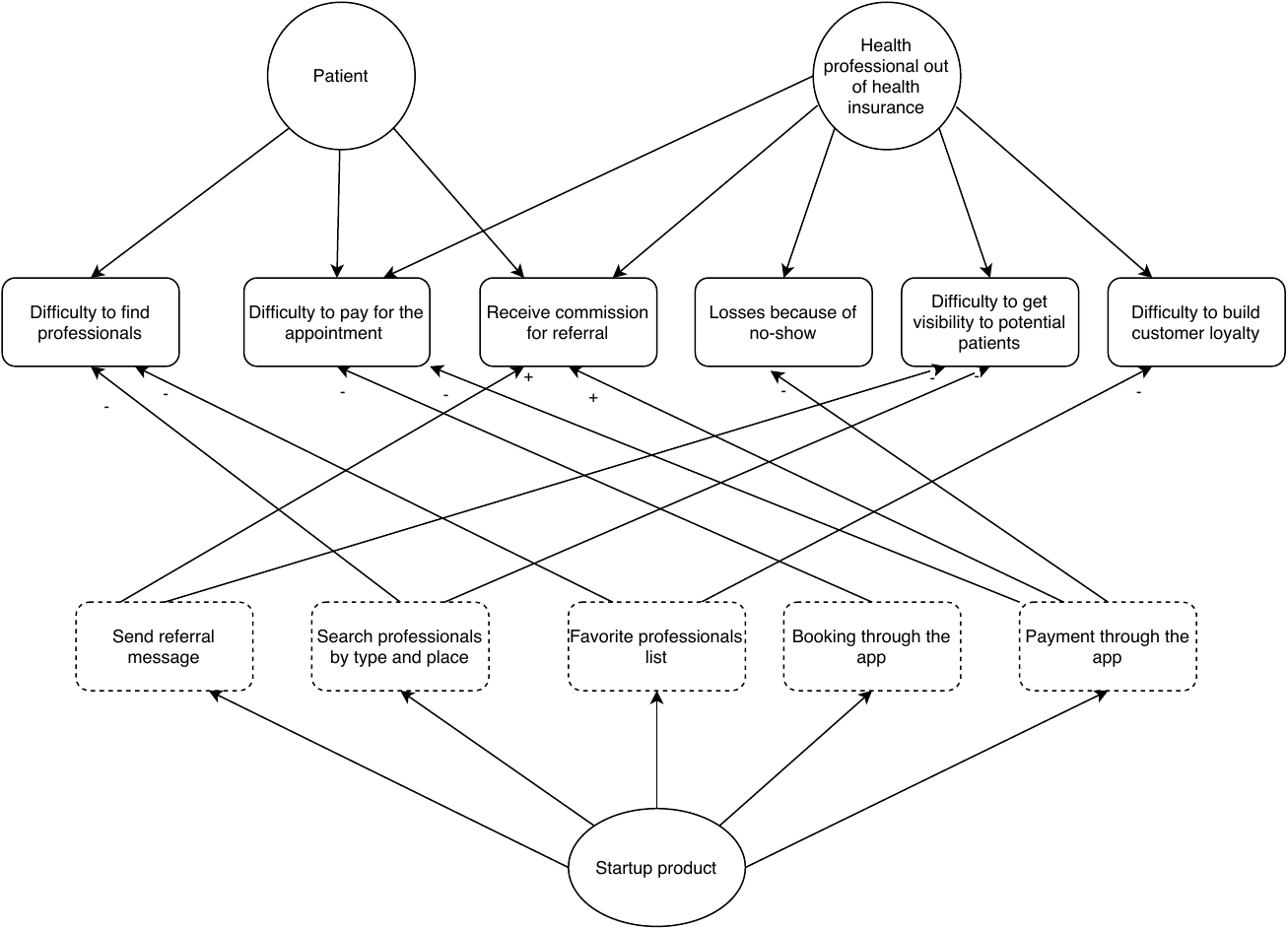}
	\caption{Cognitive map created for startup F.}
	\label{fig:case_f_cognitive_map}
\end{figure*}

Case F was a Brazilian startup developing an app to connect patients to health professionals generally not found through insurance companies like psychologists, nutritionists, and chiropractics. By the time of data collection, the company had developed most of the software solution and was planning to launch it shortly. Therefore, the startup was at the end of the inception stage. The founder team consisted of two people: one concentrated on software development and the other on product conception and other issues. We performed interviews with the latter. The result of the first interview is the cognitive map depicted in Fig.~\ref{fig:case_f_cognitive_map}.    

Based on the cognitive map, we identified 23 hypotheses of which eight were related to problems (e.g., ``the patient has difficulty to find professionals''), ten to value (e.g., ``searching professionals by type and place decreases the difficulty to find professionals''), and five to the product (e.g., ``the team developing the product is capable of implement the search for professionals by type and place''). 

When we asked the interviewee to rate the problem hypotheses, he claimed to have validated seven out of the eight hypotheses based on his own experience using those services or talking to professionals they know. The founder regarded five validated as representing a high risk to the business if not valid, one representing medium risk, and two, low. The hypothesis not validated was about the referral program. The founder attributed this classification to the fact that this feature was not essential to product viability. Regarding the value hypotheses, he acknowledged that they had not evaluated them, but it will be possible to evaluate them as soon as they have launched the product. Regarding the risk, five were considered high, one was classified as medium, and four with low risk. Finally, since the product was almost ready, the product hypotheses had already been validated, and three of them had high risk, while two were low risk.

Regarding the feedback, the founder deemed the process useful because he became ``more conscious about patients and health professionals' problems and, especially, the value hypotheses, that were not under [his] radar before.'' He found the diagram and the process easy to use because of the relationship with the hypotheses types. According to him, the diagram and process are clear. He claimed to identify two new problem hypotheses and three value hypotheses. Nevertheless, the founder observed that the process did not identify a potential hypothesis: patients have difficulty booking appointments with professionals. He mentioned that this aspect became clear to him while discussing with the facilitator after the diagram elicitation process ended.

\subsubsection{Case G}

\begin{figure}
	\centering
	\includegraphics[width=\columnwidth]{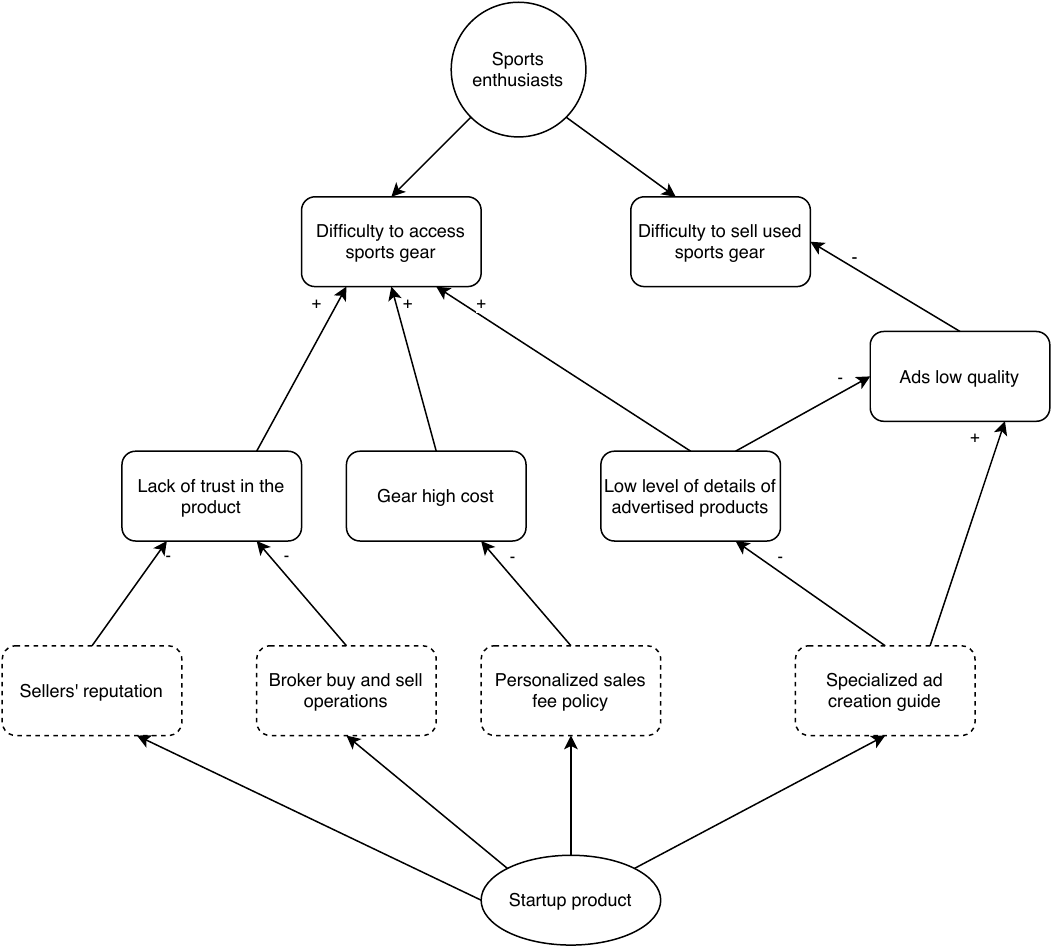}
	\caption{Cognitive map created for startup G.}
	\label{fig:case_g_cognitive_map}
\end{figure}

Case G was another Brazilian startup developing an online marketplace for second-hand sports gear. By the time of data collection, the startup had been created for a year, and the service was already online. Therefore, we classified this startup in the stabilization stage. We ran a session with the startup founder that led to the cognitive map depicted in Fig.~\ref{fig:case_g_cognitive_map}. 

Based on this map, we generated 16 hypotheses from which two were related to the problems (e.g., ``sports enthusiasts have difficulty to access sports gear''), ten to value (e.g., ``the lack of trust in the products increases the difficulty to access sports gear''), and four to the product (e.g., ``the team implementing the solution is capable of implementing seller's reputation''). 

The founder answered a questionnaire about his evaluation of generated hypotheses. Regarding the two problem hypotheses, the founder answered that they represent a high risk to the business if not valid, but he claimed to have validated them based on ``field and online surveys.'' The founder had a similar view for the product hypotheses: all represented high risk if not valid but were validated based on the actual implementation. Concerning the value hypotheses, out of the 11 hypotheses, the founder considered six validated and five not. The validation came either from ``field and online surveys'' or from the service's current users. The founder evaluated the risk as high for one, medium for four, and low for one. For those not validated, two were high risk and three medium.

Regarding the feedback, the founder considered the process useful given ``the importance of the validation process to every startup.'' He deemed the diagram and process easy to use and clear. He observed that ``there were crucial points that were unnoticed [before]'' but he claimed that this happened because the product was still not ready.

\subsubsection{Cross-case analysis}

In this section, we compare the results obtained from the different startups. Through this comparison, we discuss the utility of HyMap across different development stages. To facilitate this analysis, Table~\ref{tab:summary} summarizes how the founders perceived the hypotheses identified: for each case, it shows how many hypotheses were identified, how the founders classify them according to the risk in case of not being valid (high, medium, or low) and if they considered that these hypotheses had already been validated.

\begin{table*}[!ht]
	\renewcommand{\arraystretch}{1.2}
	\caption{Summary of hypotheses obtained by case. The letters L, M, and H stand for the risk level perceived by the founders: low, medium, and high.}
	\label{tab:summary}
	\begin{tabular}{ccccccccccccccc}
		\hline
		\multirow{2}{*}{Case} & \multirow{2}{*}{Stage}             & \multirow{2}{2cm}{Status as perceived by the founder} & \multicolumn{4}{c}{Problem} & \multicolumn{4}{c}{Value} & \multicolumn{4}{c}{Product} \\
		&                                    &                             & L    & M    & H   & Total   & L   & M   & H   & Total   & L    & M    & H   & Total   \\[10pt] \hline
		\multirow{2}{*}{E}    & \multirow{2}{*}{Inception (begin)} & Validated                   & -    & 3    & 1   & 4       & -   & 3   & 8   & 11      & -    & -    & -   & -       \\
		&                                    & Not validated               & -    & -    & -   & -       & -   & 1   & -   & 1       & 1    & 1    & 4   & 6       \\ \hline
		\multirow{2}{*}{F}    & \multirow{2}{*}{Inception (end)}   & Validated                   & 2    & -    & 5   & 7       & -   & -   & -   & -       & 2    & -    & 3   & 5       \\
		&                                    & Not validated               & -    & 1    & -   & 1       & 4   & 1   & 5   & 10      & -    & -    & -   & -       \\ \hline
		\multirow{2}{*}{G}    & \multirow{2}{*}{Stabilization}     & Validated                   & -    & -    & 2   & 2       & 1   & 4   & 1   & 6       & -    & -    & 4   & 4       \\
		&                                    & Not validated               & -    & -    & -   & -       & -   & 3   & 1   & 4       & -    & -    & -   & -       \\ \hline
	\end{tabular}
\end{table*}

Comparing the different cases, we could observe similar results. Regarding problem hypotheses, founders claimed that, although having high risk, they had validated these statements. The exception was two problem hypotheses for case F that were regarded as not validated but with a minor risk since they were related to a feature not essential to the product. Founders claimed to have validated these hypotheses, mainly based on their own experiences or interaction with customers from the targeted market.

For value hypotheses, we obtained different results. The founder of E claimed that most of the hypotheses were validated based on her experience and surveys with potential customers. However, for cases F and G, founders admitted that some hypotheses have not being validated although the product is ready and, in case G, already online. For startup G, the one in the stabilization stage, the founder claimed that some hypotheses were validated by the product usage. The fact that startup F already had its product ready but had not validated all the value hypotheses for its product means that it had the risk of not being useful to customers and not being adopted. It means that the product launch will be the experiment to test these hypotheses. Similarly, startup G used the product to evaluate many hypotheses. If these hypotheses had been identified earlier in the development process, the team could have tested them with cheaper experiments rather than building the whole product. Therefore, it is probably more useful to run HyMap in the startup earliest stages to avoid unnecessary investments.

A common aspect regarding the two types of hypotheses mentioned above is the prevalence of founders' previous experience to support these statements' validity. Although the interviewees' claim of validity, there was no systematic approach to evaluate these hypotheses and, consequently, a risk that they are not valid, leading to the development of unneeded solutions.

Regarding product hypotheses, the founders considered these hypotheses validated for the two cases (F and G) where the initial product was ready. For case E, since the product development had not started, the founder believes that the hypotheses had not been validated. In all cases, founders generally considered the hypotheses as high risk to the product viability.

Regarding the feedback about the technique, all founders answered that it allowed them to see their business idea better. Although not highlighting unnoticed elements, founders claimed that the practice gave a structured form to the product idea. 

\section{Discussion}
\label{sec:discussion}

With HyMap, we aimed to answer our research question: ``How can hypotheses be systematically defined in early-stage software startups to support experimentation?'' To verify if the artifact reached this goal, we proposed analyzing the criteria: utility, quality, and effectiveness. 

Regarding utility, the technique used in the startups as described in Section~\ref{sec:evalution} demonstrated its capability of eliciting hypotheses even though, as more mature the startup becomes, the higher the probability that teams have already confirmed the hypotheses. Nevertheless, the technique identified hypotheses not validated, even for the startup on the stabilization stage. Besides that, all founders mentioned the value of having a graphical overview of their business. For instance, such visualization may help communicate product aspects to other stakeholders like a marketing agency. We can also expect that teams could use the map as a living document updated according to how the startup progress and validates or updates hypotheses about its product, customers, and market.

Concerning quality, we evaluated three attributes: ease of use, clarity, and self-containedness. The small number of visual language elements and the process simplicity are good indicators for ease of use. The amount of time spent creating the map (around one hour in each case) demonstrates that the technique demands few resources, an essential aspect in the startup resource and time-constrained context. This aspect is related to the self-containedness that we observed by the facility with which the facilitator grasped the instructions, how the evaluation sessions ran, and the results they reached. Finally, the diagram and the process are perceived as clear, as shown by the maps displayed, the process description, and the founders' feedback. Future work could investigate how to further improve the visual notation, following aspects as the graphic economy and visual expressiveness, as suggested by Moody~\cite{Moody2009}. An aspect that we have not explicitly evaluated was if the technique could reach a complete set of hypotheses, that is, if it could identify all of them. Based on the example of case F, it is clear that such an aspect was not reached. This fact is probably related to our choice of using a freehand approach rather than pairwise judgments of causal relationships that are linked to a better coverage~\cite{Hodgkinson2004} as we discussed in Section~\ref{ssec:cognitive_mapping}. Besides that, our goal was to reach an initial set of hypotheses that could be extended and iteratively refined throughout the startup existence. In this regard, hypotheses behave similarly to requirements in a software project, that is, they keep changing during development~\cite{Williams2003}.

Effectiveness is the most challenging aspect to evaluate. Given that founders were not used to thinking about hypotheses in startups, a proper explanation of the concept would involve examples that might bias the evaluation results. Therefore, we stopped asking at the beginning of the evaluation session what their hypotheses were. Then, to support the claim of an effective technique, we should count on founders' feedback during the whole process of artifact construction and the lack of validation of hypotheses observed even in later stages. Regarding the latter, our results showed that startups, even with an initial version of the product already available to customers, the product still led to hypotheses without proper backing. For instance, in Case F, founders have not validated that ``the search for professionals decreases the difficulty to find professionals.'' In Case G, the same happened to the hypothesis: ``the lack of trust in the products increases the difficulty to obtain sports gear.''

The results of our evaluation show that HyMap have several of the expected attributes, i.e., ease to use, usefulness, clarity, and self-containdness. However, other studies could better evaluate other aspects, such as effectiveness. Nevertheless, we agree with Hevner~\cite{Hevner2013} who writes: ``When a researcher has expended significant effort in developing an artifact in a project, often with much formative testing, the summative (final) testing should not necessarily be expected to be as full or as in-depth as evaluation in a behavioral research project where the artifact was developed by someone else.'' As future work, we suggest comparing HyMap with different approaches such as Assumption Mapping~\cite{Bland2019}.

An interesting consequence of the diagram developed is a relationship among hypotheses and requirements. This result is in line with the concept of Dual-track development~\cite{Miaskiewicz2017,Sedano2020}, a reconciliation between human-centered design and agile methods. According to Sedano et al.~\cite{Sedano2020}, ``a software project comprises two continuous, ongoing, parallel tracks of work'' where one generates feature ideas and the other uses these ideas to build the product. However, HyMap shows an opposite flow: potential requirements leading to guidance to better understand the product. For instance, requirements imagined by the founders could guide the proposal of A/B tests~\cite{Ros2018} that will inform the team of what to implement.

\begin{figure}
	\centering
	\includegraphics[width=\columnwidth]{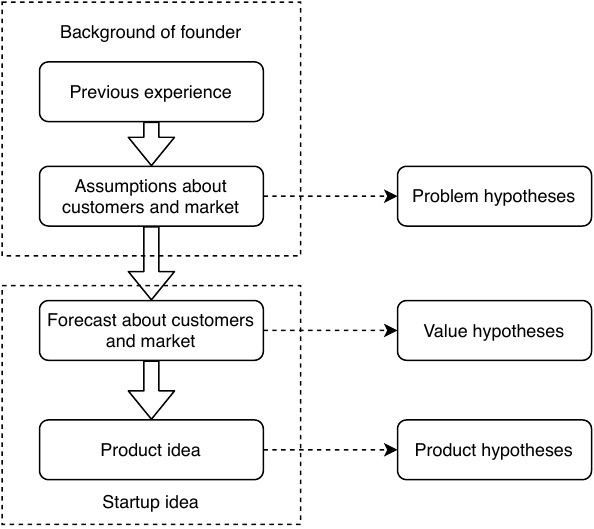}
	\caption{The relationship between the idea creation process and the hypotheses types identified by the HyMap process.}
	\label{fig:hypotheses_types}
\end{figure}

Still, regarding the hypotheses types, we can map them to the steps in the idea creation process, depicted in Fig.~\ref{fig:idea_creation}, as shown in Fig.~\ref{fig:hypotheses_types}. Based on the previous personal experience, the founder builds an understanding of the target market and customers, leading to problem hypotheses through the HyMap technique. Based on this understanding, the founder forecasts how the customers and market will behave. The elicitation process extracts these assumptions as value hypotheses. Finally, the product envisioned by the founder that would carry out her expectations leads to the product hypotheses relating to the solution feasibility and the team capability of doing it.

Besides that, the frequency of their own experience as an answer to how the hypotheses were validated is another piece of evidence to support the idea creation process. It was also evident in these cases that founders based their product ideas on their previous experiences or observations of the target market and not necessarily with proper backing. These weak validations are another evidence to support the importance of identifying hypotheses earlier in the startup development. It would also be essential to improve the validation mechanisms in startups, but this goal is beyond this paper's scope. Nevertheless, this study is a step towards this goal.

Another interesting aspect of hypothesis types, problem, value, and product, is that they can act as elements to the prioritization activity, an essential aspect of hypotheses engineering~\cite{Melegati2019}. For instance, if customers do not experience the problems the product is aiming to solve, represented by problem hypotheses, it is hard for the product to succeed. Therefore, we could expect that these are probably the first to be evaluated. Then, the value hypotheses represent assumptions if the proposed solution would eliminate or reduce the problems. Hence, they are the next to be evaluated. Finally, the startup could evaluate if the team is technically able to execute the proposed solution.

Our results are also related to the three approaches to software development identified by Bosch et al.~\cite{Bosch2018}. The first approach is the conventional requirement-driven. The second is the outcome or data-driven, that is essentially experimentation, where teams use data to improve and optimize the system. Finally, the third approach is the rising AI-driven software development, where the software would be automatically updated by machine learning algorithms trained with user data. The authors argue that these approaches co-exist and should be used according to the needs. The relationship among potential features and hypotheses identified in HyMap diagrams suggests an intertwined process where requirements also could take to hypotheses.

\subsection{Contributions}

Our study presents contributions to theory and practice. Regarding theory, the first phase results showed that founders develop an understanding of the customers and markets based on their previous experience and use this perception to develop the idea and forecast how it will behave with customers. The artifact evaluation also corroborated this result. Based on the final artifact, it was possible to observe at least three different types of hypotheses: product, value, and problem. In an analysis of practitioners' literature~\cite{Melegati2021}, we had identified these types, along with customer and market-related hypothesis types. Therefore, this study corroborates this result and connects different types of hypotheses with the steps in the idea formation process. Finally, the linkage between requirements and hypotheses shown in the cognitive maps built in HyMap corroborates the existence of a relationship between requirements and hypotheses. These results contribute to understanding how founders have initial ideas for their software startup product and how new research could further improve Hypotheses Engineering and its activities.

Concerning contributions to practice, HyMap is the primary result: a process to elicit hypotheses in software startups based on a visual language used to depict a cognitive map representing how founders understand their customers and market, and a systematic set of steps to create this map. Early-stage software startups could use this technique to guide their initial steps more systematically. Other stakeholders like accelerators or investors could apply this technique to the companies to which they are related. The tool could also be used in educational environments, i.e., graduate or undergraduate courses on digital entrepreneurship, to highlight the underlying assumptions founders have when creating their products. We envision that this practice could be useful for other software development teams when creating new features for consolidated products, but investigating this suggestion is beyond this paper's scope.

\subsection{Threats to validity}

Given that we employed multiple-case studies in two cycles of the design phase (as already described in our previous paper~\cite{Melegati2020}) and in the evaluation of the final artifact, we deemed it essential to discuss threats to the validity of these investigations. We followed the definitions given by Runeson and Host~\cite{Runeson2012}. The authors describe four aspects of validity for a case study: construct validity, internal validity, external validity, and reliability.

Construct validity concerns to what extent the case elements studied represent what the researchers have in mind. A common threat when using interviews is if the interviewee has the same understanding of terms and concepts used in the questions as the interviewer. Since the interview guide for the first phase of artifact construction focused on the business model description and evolution, such a threat is minimal. Besides that, the triangulation of data with a different team member interview decreased the threat even more. In the evaluation phase, we assessed several constructs, such as quality and utility. In this regard, we followed Yin~\cite{Yin2003} who suggests to select the changes to be studied related to the study's objectives and to demonstrate that the selected measures indeed reflect the selected changes.

Triangulation was also essential to mitigate threats to internal validity. This aspect relates to causal inferences when the researchers attribute the cause of an effect to a phenomenon, but, in reality, it is caused by a third one not considered in the analysis. In addition to triangulation, we employed peer debriefing; that is, all authors discussed the results.

External validity reflects the extent to which the results can be generalized or interesting to other people outside the studied case~\cite{Runeson2012}. As mentioned by Runeson et al.~\cite{Runeson2012}, in case studies, it is not possible to draw statistical significance. Still, the goal should be an analytical generalization of the results to cases with similar characteristics. As we argued before, the studied cases are typical software startups where the founder is the main innovation owner and dictates the requirements. Besides that, these companies generally focus on developing a solution instead of understanding the customer~\cite{Giardino2014,Gutbrod2017}. Therefore, we expect that our results are valuable to describe a large portion of early-stage software startups. Another threat to this aspect of validity was the use of only one facilitator in the evaluation. We mitigated this issue by employing a person with no previous knowledge of startups and giving a detailed, pre-defined set of steps to be followed. However, we acknowledge that personal characteristics of the facilitator might influence the results. Future work could focus on removing this figure, for instance, by using an integrated system to make the questions and draw the map.

Finally, reliability concerns to what extent the results are dependent on the researchers that performed the study. That is, if another researcher conducts the same study, she will reach similar conclusions. To improve this aspect, we described all the steps for data collection and analysis in all artifact construction cycles and the evaluation.

\section{Conclusions}
\label{sec:conclusions}

Experimentation is a useful approach to guide software development in startups. However, the lack of defined practices to guide these teams is a reason for the reduced use of this approach. Given that identifying hypotheses is the first step to create experiments, this study focused on developing a practice to perform this task in early-stage software startups. Following a Design Science Research approach, we performed three phases to build a process to systematically elicit hypotheses for software startups based on a visual language to depict cognitive maps of startup founders and a systematic process to extract them. We evaluated these artifacts on three startups in early-stages of development.

Besides the prescriptive contribution represented by HyMap, this study presents a theoretical contribution by describing the process of idea formation from founders of these companies. Besides that, we mapped different types of hypotheses to distinct steps in the idea formation process.

As mentioned earlier, a more extensive evaluation of HyMap would be valuable future work, probably using controlled experiments or longitudinal case studies. For instance, a startup could use HyMap in an early-stage and keep the map as a live document, changing it according to how the team improves their understanding of the customers and market. Another interesting work would be to assess the technique outside the startup context, for instance, when adding new features to market-driven products where development teams create software for a market of users rather than specific customers. Besides that, other studies could improve the completeness of the hypotheses set generated by the technique, probably extending the language and the process. These enhanced processes could also tackle other types of hypotheses. 

\bibliography{refs}

\end{document}